# An Efficient Mixture of Deep and Machine Learning Models for COVID-19 and Tuberculosis Detection Using X-Ray Images in Resource Limited Settings

Ali H. Al-Timemy[1], Rami N. Khushaba[2], Zahraa M. Mosa[3] and Javier Escudero[4*]

[1] Biomedical engineering department, Al-Khwarizmi College of Engineering, University of Baghdad, Jaderiyah, Baghdad, Iraq. Email: ali.altimemy@kecbu.uobaghdad.edu.iq

[2] Australian Centre for Field Robotics, the University of Sydney, 8 Little Queen street, Chippendale, NSW 2008, Australia. Email: Rami.Khushaba@sydney.edu.au

[3] Department of Physics, College of Science for Women, University of Baghdad, Baghdad, Iraq. Email: zahraronan_1984@yahoo.com

[4] School of Engineering, Institute for Digital Communications, The University of Edinburgh, Edinburgh, King's Buildings, EH9 3JL, UK.

[*] *Corresponding author:* javier.escudero@ed.ac.uk

## Abstract

Clinicians in the frontline need to assess quickly whether a patient with symptoms indeed has COVID-19 or not. The difficulty of this task is exacerbated in low resource settings that may not have access to biotechnology tests. Furthermore, Tuberculosis (TB) remains a major health problem in several low- and middle-income countries and its common symptoms include fever, cough and tiredness, similarly to COVID-19. In order to help in the detection of COVID-19, we propose the extraction of deep features (DF) from chest X-ray images, a technology available in most hospitals, and their subsequent classification using machine learning methods that do not require large computational resources. We compiled a five-class dataset of X-ray chest images including a balanced number of COVID-19, viral pneumonia, bacterial pneumonia, TB, and healthy cases. We compared the performance of pipelines combining 14 individual state-of-the-art pre-trained deep networks for DF extraction with traditional machine learning classifiers. A pipeline consisting of ResNet-50 for DF computation and ensemble of subspace discriminant classifier was the best performer in the classification of the five classes, achieving a detection accuracy of 91.6± 2.6% (accuracy ± Confidence Interval (CI) at 95% confidence level). Furthermore, the same pipeline achieved accuracies of 98.6±1.4% and 99.9±0.5% (± CI) in simpler three-class and two-class classification problems focused on distinguishing COVID-19, TB and healthy cases; and COVID-19 and healthy images, respectively. The pipeline was computationally efficient requiring just 0.19 second to extract DF per X-ray image and 2 minutes for training a traditional classifier with more than 2000 images on a CPU machine. The results suggest the potential benefits of using our pipeline in the detection of COVID-19, particularly in resource-limited settings as it relies in accessible X-rays and it can run with limited computational resources. The final constructed dataset named COVID-19 five-class balanced dataset is available from: https://drive.google.com/drive/folders/1toMymyHTy0DR_fyE7hjO3LSBGWtVoPNf?usp=sharing.

**Keywords:** COVID-19, Deep features; Machine learning; Pneumonia; ResNet-50; Tuberculosis.

# 1. Introduction

During December 2019, the world witnessed the outbreak of a new, previously unknown, virus and disease in Wuhan, China, resulting in an infectious disease known as Coronavirus Disease COVID-19[1], which is caused by a recently discovered coronavirus. Coronaviruses are a large family of viruses which may cause respiratory infections ranging from the common cold to more severe diseases such as Middle East Respiratory Syndrome (MERS) and Severe Acute Respiratory Syndrome (SARS). The most common symptoms of COVID-19 include fever, cough, difficulty in breathing, and tiredness, which may develop to pneumonia and potentially leading to death [1,2]. The symptoms of pulmonary diseases may be confused and there might be a misdiagnosis of COVID-19 cases with that of other pulmonary diseases, such as other types of viral pneumonia, bacterial pneumonia, and Tuberculosis (TB). This complicates the doctor's task to detect COVID-19 quickly to manage the disease, isolate the patient, and trace recent contacts if relevant.

In the current situation of the COVID-19 crisis, there are more than 12 million cases around the world and the numbers are rising at a rate of new 200000 cases per day[2]. Multiple countries are experiencing an increase in the number of patients admitted to hospital, which may compromise their healthcare systems. The situation is even more precarious in low-income countries that may have difficulties in accessing some of the most recent testing technology.

The current testing utilized to confirm COVID-19 cases is the Reverse Transcription Polymerase Chain Reaction (RT-PCR) [3]. The accuracy of the test is around 60-70% and it takes few hours to 2 days to get the results [4]. Other tests recommended by the World Health Organization (WHO), but not in mass use yet, include the Cobas SARS-CoV-2 Qualitative assay[3]. Due to the high number of cases affected every day, the waiting time to obtain the results may be longer. Furthermore, tests are difficult to access, expensive and are unavailable to certain hospitals or regions.

TB is a very serious and deadly pulmonary disease. In 2018 alone, 10 million people had TB worldwide and it caused 1.5 million deaths. It is regarded as a major cause of death from a single infectious agent worldwide [5]. Low-middle income countries account for 2/3 of the TB cases, including India which has the largest number of TB cases, followed by China, Indonesia, Philippines, Pakistan, Nigeria and

---

[1] https://www.who.int/emergencies/diseases/novel-coronavirus-2019/question-and-answers-hub/q-a-detail/q-a-coronaviruses#:~:text=symptoms
[2] https://www.worldometers.info/coronavirus/
[3] https://www.who.int/news-room/detail/07-04-2020-who-lists-two-covid-19-tests-for-emergency-use

Bangladesh [5]. Multidrug Resistant (MDR) TB, a severe type of TB, which requires chemotherapy and expensive toxic medications for treatment, stays a crisis and threat for public health, where its main load occurs in 3 countries including India, China and the Russian Federation. In the current COVID-19 pandemic, TB is an issue specially. Both TB and COVID-19 spread by close contact between people and the symptoms may be similar to COVID-19, such as fever, cough, tiredness [6]. Thus, low-income countries may face a health crisis where the COVID-19 crisis is compounded by the presence of TB. Large number of COVID-19 cases[4] are reported in India (766000 cases), Pakistan (230000 cases) and Bangladesh (170000), which results in less attention given to TB. In such cases, there is a need to achieve fast TB and COVID-19 detection. The COVID-19 pandemic may cause a global reduction of 25% in the detection of TB. This could lead to an increase of 13% increase in TB deaths due to disruption to TB detection and treatment [7].

Radiology can be utilized to help in the detection of COVID-19 in images of Computerized Tomography (CT) [3,8–10] or X-rays [4,11,12] since darkened spots and also opacities that looks like ground glass, can be observed [3]. Thus, it is suggested that X-ray and CT scans can help to select and quantify COVID-19 and aid doctors to detect this disease quickly [12,13]. CT is relatively complex modality and it is not available in all clinics. On the other hand, an X-ray scan is a simple modality and it can be available in all hospitals and in small clinics. Hence, it has been proposed for Computer Aided Diagnosis (CAD) for aiding COVID-19 detection [14–17] and TB screening [18,19].

In the healthcare field, a new subbranch of artificial intelligence algorithms denoted as Deep Learning (DL) is being increasingly developed as a CAD tool to help doctors achieve better clinical decisions with high precision [20]. Therefore, achieving good performance on disease detection problems. Similarly, DL methods can be an aid to the doctors for improving the quality of the detection of COVID-19 [4,11,12,21] and TB [18,19]. Moreover, automatic screening of TB has been proposed with deep learning [18,19,22]**.** However, deep learning requires GPU, which is not a privilege for limited resource settings.

---

[4] https://www.worldometers.info/coronavirus/

A convolutional neural network (CNN), one of the DL methods, has been applied to the medical field [23–25] to tackle a variety of problems such as COVID-19 detection with CT and X-ray images [8–10,14,15,17,26–28]. Several new CNN designs have been proposed for COVID-19 detection such as COVID-Net [21], CovidxNet [14], CoroNet [12], DarkCovidNet [4], COVIDiagnosis-Net [29], and nCOVNet [30]. Despite the remarkable research, the main challenges with CNN includes its need for a large number of images for training the network in addition to the long training time, even with GPU support.

On the other hand, Transfer Learning (TL) is proposed to deal with the challenge of the need for a large set of images and long training times of CNN. A pretrained CNN such as ResNet-50 [17], VGG-19 [17], DenseNet-201 [31] or MobileNet-v2 [32] can be utilized to learn a new task by fine tuning of the last fully connected layers. The pretrained networks are trained on a dataset of ImageNet [33] which has more than million images with 1000 classes. Despite its simplicity, this approach still requires long training time. TL has been applied to COVID-19 detection from X-ray images [4,12,14–18,21,22,26–28,35–38] where pretrained networks such as ResNet-50 [26,35,39], VGG-19 [16,40], DenseNet-201 [41] and Xception [39] has been used for COVID-19 detection with X-ray images.

Inspiring research has been done to tackle the COVID-19 detection with X-ray images. Panwar *et al.* proposed nCOVNet [30] for COVID-19 detection for 2-class dataset (142 normal and 142 COVID-19 images). A detection accuracy of 88% was obtained with 70%/30% training and testing split. Wang and Wong [21] developed COVID-Net and validated it on a dataset of 358 COVID-19, 5,538 normal and 8,066 pneumonia images with 91% sensitivity for COVID-19 cases with 70%/30% training and testing data split, respectively. It should be noted that the 3-class images were not balanced. DarkCovidNet was proposed in [4] to detect COVID-19 for 3-class dataset including 125 COVID-19 cases, 500 pneumonia and 500 normal images with an accuracy of 87.2%, utilizing the 5-fold cross validation to avoid overfitting. However, only 125 cases were included in their model. Khan *et al.* [12] proposed CoroNet CNN to detect COVID-19 from X-ray images of 4-class dataset of 284 COVID-19, 310 normal 330 pneumonia bacterial and 327 pneumonia viral. Their dataset represents the largest balanced 4-class COVID-19 dataset so far. Detection accuracy of 89.6% was obtained with 4-fold cross validation with Google Collaboratory Ubuntu server and Tesla K80 graphics card.

Deep Features (DF) extraction is a process in which the features are directly acquired from the last fully connected layer of a pretrained deep learning network e.g., AlexNet, ResNet-50 and VGG-19, without the need for re-training the networks. The main advantage of employing a pretrained model is huge savings in the computational time required to train these models from scratch, and hence the possibility of running these models on a CPU machine without the need for expensive GPU enabled computational powers. An initial attempt by Sethy *et al.* [35] utilized 127 images of 3-class of COVID-19, pneumonia (viral and bacterial) and normal, 80%-20% data split of training and testing was used to evaluate the model. An accuracy of 95.33% was obtained with ResNet-50 deep features and Support Vector Machines (SVM) classifier. However, it is not known how the performance of ResNet-50 compares to other pre-trained DL models, especially given that the dataset contained a small number of images. In [36], ResNet1523 and XGBoost classifiers were evaluated on a 3-class dataset (normal 1341 images, pneumonia 1345 images and only 62 COVID-19) with 30% holdout. From previous two studies, DF have been proposed for COVID-19 detection, but it was not investigated in depth.

The main challenges with the previous literature are either related to: 1) the small number of COVID-19 images, in comparison to other diseases considered, included in the datasets thus leading to unbalanced datasets and/or 2) the need for GPU resources to train the newly proposed CNN or the pretrained CNNs. Moreover, the class separability of COVID-19 DF is not known in problems where a larger number of diseases is considered. 3) The severe case of COVID-19 and TB is also not previously investigated, to the best of our knowledge.

In this paper, we harvest the power of the good feature presentation of 14 state-of-the-art pretrained deep networks and the simplicity of machine learning classifiers for COVID-19 detection on a CPU machine in just under ten minutes for the whole training and testing, something that makes the proposed CAD suitable for low computational power settings. In addition, the proposed DF pipeline is evaluated on a balanced 2186 X-ray images of 5 classes including COVID-19, normal, bacterial pneumonia, viral pneumonia and TB.

The main contributions of the current study are: 1) We constructed a five-class COVID 19 dataset, named COVID-19 five-class balanced dataset, including a large number of COVID-19 and tuberculosis images, which was not investigated before to the best of our knowledge. 2) A pipeline of features from

14 individual state-of-the-art pretrained deep networks combined with machine learning classifier are investigated using a five-fold cross validation scheme to avoid overfitting, without the need to train the pretrained networks. 3) The proposed pipeline can run on a CPU machine, which makes it simple and efficient, and suitable for low-middle income countries. 4) Five-class separability analysis of COVID-19 DF using Silhouette criterion clustering evaluation and high dimensional t-SNE visualization were investigated to understand the separation of the 2186 5-class images.

## 2. Methodology

### 2.1 COVID-19 5-Class Balanced Dataset

The 5-class COVID-19 dataset, constructed in this study, consists of 3 main sources. The COVID-19 cases are acquired from Cohen *et. al* dataset [37] which is collection of X-ray and CT images of variety of pulmonary diseases, including COVID-19, SARS and MEARS, compiled from many sources. The dataset is updated on a regular basis to include more cases. Until 15/6/2020[5], it had 752 X-ray images, among them 435 COVID-19 X-ray images which were utilized in this study. This dataset has the largest number of X-ray images of COVID-19 cases. The CT and lateral X-ray images were excluded from this study. The average age of the COVID-19 patients is 54.6 years ± 16.7 (mean ± standard deviation), including 256 males and 136 females, and the gender for 43 images was not provided, since the metadata accompanying the dataset is not complete.

The second source of images is the Pneumonia and normal dataset [38] which has 5863 X-ray images[6] of 3-class including normal, pneumonia bacterial and pneumonia viral. We randomly selected 439 images of each class from the pneumonia dataset to be included in the study, to construct the COVID-19 balanced 5-class dataset.

The last source of X-ray images is TB dataset from U.S. National Library of Medicine, consisting of two datasets of chest X-ray. TB dataset were made available to expedite research in CAD of pulmonary diseases specially TB[7] [18,22]. The TB dataset includes 394 images from 2 sources (336 images from China set and 58 images from Montgomery County set) [18,22]. The TB X-ray images has a slightly less number than other classes in this study which may have an influence on the detection performance due to unbalanced classes [42]. To balance the TB class, augmentation [43] of random images has been done by rescaling 40 random X-ray images. The details of the constructed COVID-19 5-class balanced dataset is illustrated in Table 1. The current COVID-19 5-class dataset has a larger number of COVID-19 X-ray images, than previous studies [4,11,12,35]. Samples of X-ray images of the 5 classes included in this study, with their labels, are shown in Fig. 1.

---

[5] https://github.com/ieee8023/covid-chestxray-dataset
[6] https://www.kaggle.com/paultimothymooney/chest-xray-pneumonia
[7] https://www.kaggle.com/kmader/pulmonary-chest-xray-abnormalities?select=ChinaSet_AllFiles

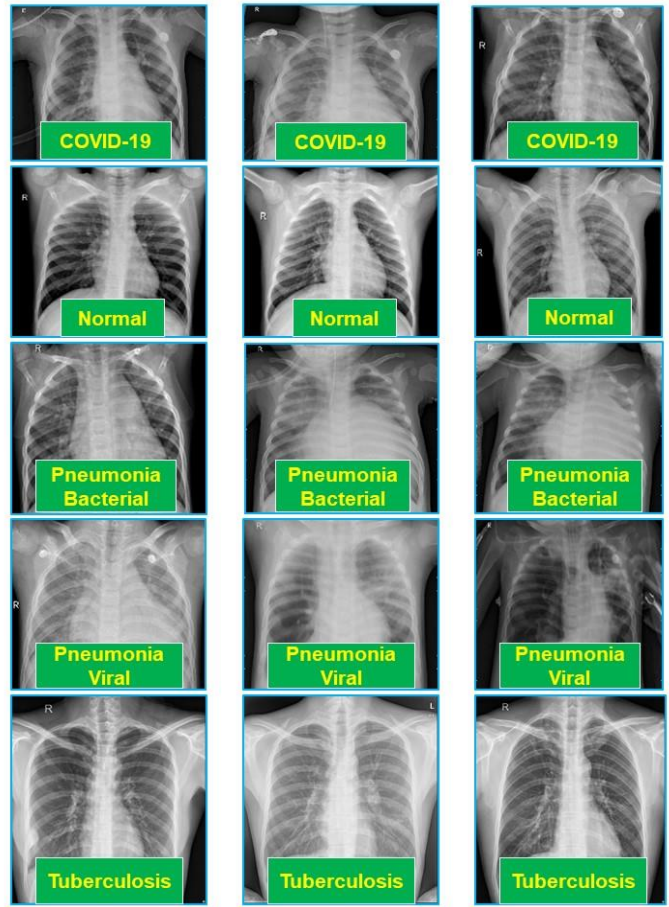

**Fig.1** Samples of the five-class X-ray images for the COVID-19, normal, pneumonia bacterial, pneumonia viral and TB.

**Table 1** The details of the COVID-19 5-classes dataset.

| Class | Number of x-ray images |
|---|---|
| **COVID-19** | 435 |
| **Normal** | 439 |
| **Pneumonia-bacterial** | 439 |
| **Pneumonia-viral** | 439 |
| **Tuberculosis** | 434 (394+40 augmented) |
| **All 5-class dataset** | 2186 |

## 2.2 The Pipeline of Deep Feature Extraction from Pretrained Networks and Machine Learning Classification

In this section, the details of the proposed COVID-19 detection method will be presented. Fig. 2 presents the block diagram of the proposed pipeline for COVID-19 detection with deep features and machine learning classifiers on 5-class COVID-19 balanced dataset. The details of each part of the methodology will be presented in the next paragraphs.

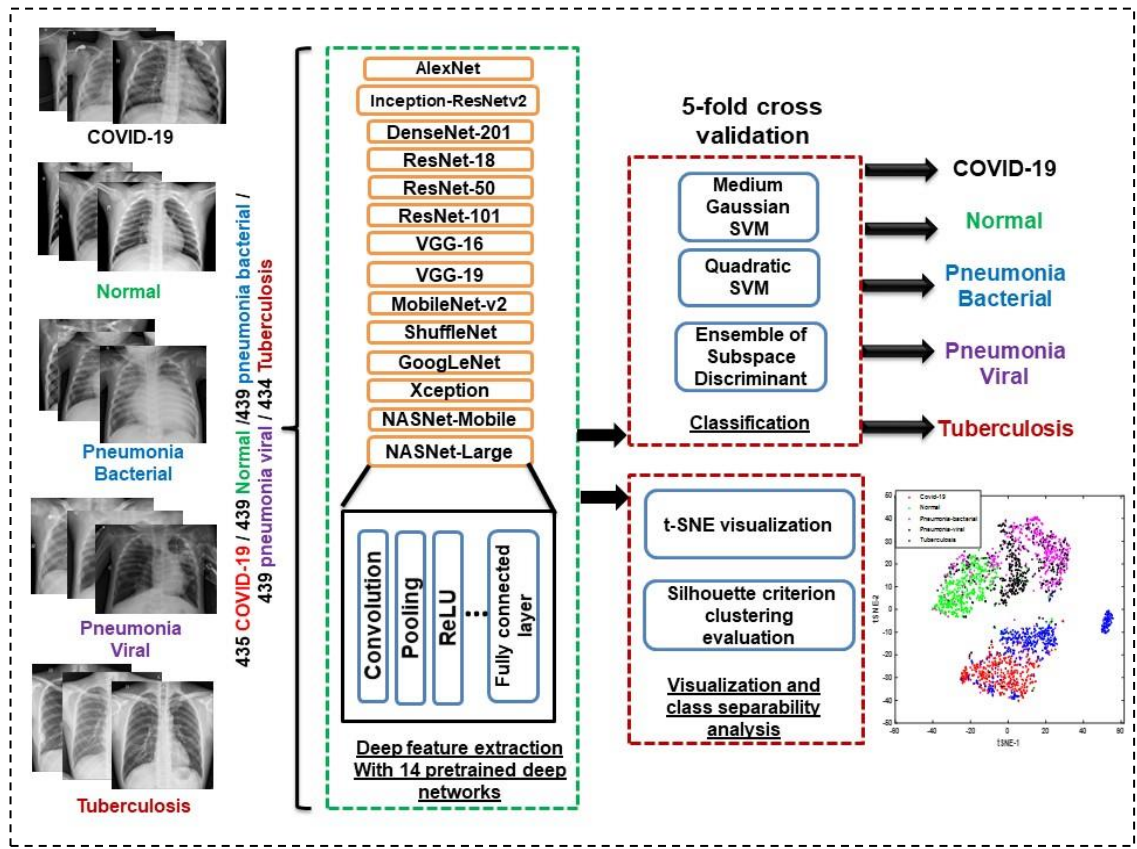

**Fig.2** The block diagram of the proposed 5-classes COVID-19 detection method.

In order to extract the deep features from the pretrained networks, 14 individual state-of-the-art pretrained networks were included in this study ranging from the networks with simpler designs such as AlexNet and ShuffleNet to the advanced complex designs such as VGG-19 and NASNet-Large which requires large GPU power for training. These networks are pretrained on ImageNet dataset [33] which has one million images for 1000 classes. In this study, we will input the X-ray images to each individual pretrained network and extract the feature vectors directly at the fully connected layer, without the need for training the network on the X-ray images. The full details of the 14 pretrained networks utilized in this study for feature extraction, with their input size, number of layers the name of the fully connected layer as well as the number of parameters are illustrated in Table 2. Preprocessing was performed to change the input size of the X-ray images to match the input of the targeted pretrained network (Table 2). The size of the extracted feature vector for each network was equal to 2186 x 1000 (number of X-ray images x number of DF), for each of the 14 networks.

**Table 2** The details of the 14 pretrained deep networks utilized in this study.

|   | Network name | Reference | Input size | Fully connected layer | No. of layers | No. of parameters in Million |
|---|---|---|---|---|---|---|
| 1 | AlexNet | Krizhevsky *et al.* [44] | 227x227x3 | fc8 | 25 | 61 |
| 2 | Inception-ResNet-v2 | Szegedy *et al.* [45] | 299x299x3 | predictions | 824 | 23.2 |
| 3 | DenseNet-201 | Huang *et al.* [31] | 224x224x3 | fc1000 | 708 | 20 |
| 4 | ResNet-18 | He *et al.* [46] | 224x224x3 | fc1000 | 71 | 11.7 |
| 5 | ResNet-50 | He *et al.* [46] | 224x224x3 | fc1000 | 177 | 25.6 |
| 6 | ResNet-101 | He *et al.* [46] | 224x224x3 | fc1000 | 347 | 44.6 |
| 7 | VGG-16 | Simonyan and Zisserman [47] | 224x224x3 | fc8 | 41 | 138 |
| 8 | VGG-19 | Simonyan and Zisserman [47] | 224x224x3 | fc8 | 47 | 144 |
| 9 | MobileNet-v2 | Sandler *et al.* [32] | 224x224x3 | Logits | 154 | 3.5 |
| 10 | ShuffleNet | Zhang *et al.* [48] | 224x224x3 | node_202 | 172 | 1.4 |
| 11 | GoogLeNet | Szegedy *et al.* [49] | 224x224x3 | loss3-classifier | 144 | 7 |
| 12 | Xception | Chollet *et al.* [50] | 299x299x3 | predictions | 170 | 22.9 |
| 13 | NASNet-Mobile | Zoph *et al.* [51] | 224x224x3 | predictions | 913 | 5.3 |
| 14 | NASNet-Large | Zoph *et al.* [51] | 331x331x3 | predictions | 1243 | 88.9 |

To perform the classification of the extracted features for each of the 14 networks, we utilized machine learning classifiers. Classification Learner (CL) App. in Matlab 2019b was used to perform the classification. We utilized five-fold Cross Validation (CV) for analysis in this study to avoid overfitting, as done in [4,12,26]. To select the best classifier for the COVID-19 5-class dataset, we carried out a pilot study with multiple classifiers in the CL App, including *k*-Nearest Neighbors (KNN), Linear Discriminant Analysis (LDA), Support Vector Machines (SVM), and variety of ensemble classifiers including boosting, bagging and subspace discriminant. The best three classifiers in the pilot study were quadratic SVM, medium Gaussian SVM and ensemble of subspace discriminant classifiers. The setting for the SVM classifiers utilized in this study are as follows 1) quadratic SVM (kernel function: quadratic, box constraint level=1, one vs. one classification); 2) medium Gaussian SVM (kernel type: Gaussian, kernel scale= 32, box constraint level=, one vs. one classification). As for the ensemble of subspace discriminant classifier, random subspace ensembles [52] of a group of discriminant analysis classifier was used. Subspace ensembles have been utilized with the following settings (learner: discriminant. learning rate =0.1, number of learners=30, subspace dimension=500).

**2.3 Performance Evaluation of the Proposed COVID-19 Detection Pipeline**

The performance of the proposed COVID-19 detection was evaluated with the following classification performance measures including accuracy, precision, recall, specificity, and *F*-score, given by

$$\text{Accuracy} = \frac{Number\ of\ correctly\ classified\ images}{Total\ number\ of\ images} \quad (1)$$

$$\text{Precision} = \frac{Sum\ of\ all\ TP}{Sum\ of\ all\ TP+FP} \quad (2)$$

$$\text{Recall} = \frac{Sum\ of\ all\ TP}{Sum\ of\ all\ TP+FN} \quad (3)$$

$$\text{Specificity} = \frac{TN}{TN+FP} \quad (4)$$

$$\text{F-score} = 2 * \frac{Precision*Recall}{Precision+Recall} \quad (5)$$

where TP (True Positive), TN (True Negative), FN (False Negative) and FP (False Positive). We also estimated the Area Under the Receiver Operating Characteristics (ROC) Curve (AUC) in the CL app. in MATLAB 2019b.

An *n*-way analysis of variance (ANOVA) statistical test was employed to investigate the effects of two factors on the mean of the accuracy results. The first factor being related to the selected DL models with 14 levels (14 models) and the second being related to the selected 3 traditional classifiers (quadratic SVM, medium Gaussian SVM, and ensemble of subspace discriminant classifiers). We have also studied other class combinations, including 3-classes (COVID-19 vs. normal vs TB) and 2 -classes (COVID-19 vs. normal) to show the robustness of the proposed method with different number of COVID-19 classes.

To investigate class separability of the COVID-19 features, we utilized *t*-Distributed Stochastic Neighbor Embedding (*t*-SNE) data visualization [53,54] to reduce dimensions of the deep features computed with the pretrained networks. This was done for the 5-class, 3-class and 2-class COVID-19 features. In addition, The Silhouette criterion – a clustering evaluation measure – [55] was used to evaluate the quality of the 5-class COVID-19 features.

## 3. Results

In this section, we will present the results of the analysis of the five-class COVID-19 balanced dataset with the proposed deep feature extraction and machine learning classification pipeline. Fig.3 shows the results of the classification accuracy with the 14 state-of-the-art deep feature learning methods integrated with three classifiers (quadratic SVM, medium Gaussian SVM and ensemble of subspace discriminant). Moreover, we also estimated the Confidence Interval (CI) for the detection accuracy calculated at 95% confidence level (Fig.3). ResNet-50 and ResNet-101 were the best DF, with all classifiers, compared to the other networks. In addition, ensemble of subspace discriminant classifier was the best on average among the other classifiers. To show the effectiveness of proposed method, we calculated the time for feature extraction of all 2186 images, with all pretrained networks (Table 3), which ranges from few minutes for simple networks (AlexNet and ResNet-50) to >30 minutes for big networks such as VGG-19 and NASNet-Large. On average, 0.09 s is needed to extract feature with AlexNet for a single image. This time was calculated on a Core *i5* CPU computer with 16 GB RAM. It is worth noting that the time to perform the classification of the whole dataset with five-fold cross-validation was equal to 0.5 minutes for SVM classifiers and 2 minutes for ensemble of subspace discriminant classifier. It is noteworthy to mention that all classification in this study was done with 5-fold CV to avoid overfitting.

Using the *n*-way ANOVA, a *p*-value < 0.001 returned by this test indicated that the average classification accuracies are significantly different across the three levels of the classifiers factor (three traditional classifiers). This was followed by a Bonferroni corrected *t*-test which further indicated that, across all DL models, significant differences existed between the quadratic SVM and ensemble of subspace discriminant classifier (*p*-value = 0.002) and that between medium Gaussian SVM and ensemble of subspace discriminant classifiers (*p*-value < 0.001), but not between the two version of SVM. Similarly, there were also statistically significant differences across the 14 levels of the DL models factor, with a *p*-value < 0.001. Additionally, the combination of ResNet-50 with the ensemble of subspace discriminant classifiers was the most significantly different from other combinations of DL models and classifiers, followed by ResNet-101 with the ensemble of subspace discriminant classifiers. However, ResNet-101 required almost double the computational time of ResNet-50, per the results from Table 3.

The pipeline consisting of ResNet-50 and ensemble of subspace discriminant classifier is the best performer among all networks and classifiers (detection accuracy 91.6 ± 2.6%, accuracy ± CI), as confirmed statically, in addition to being efficient (6.9 minutes for DF and 2 minutes for classification, see Table 3). It will be chosen to perform the subsequent analysis in this study.

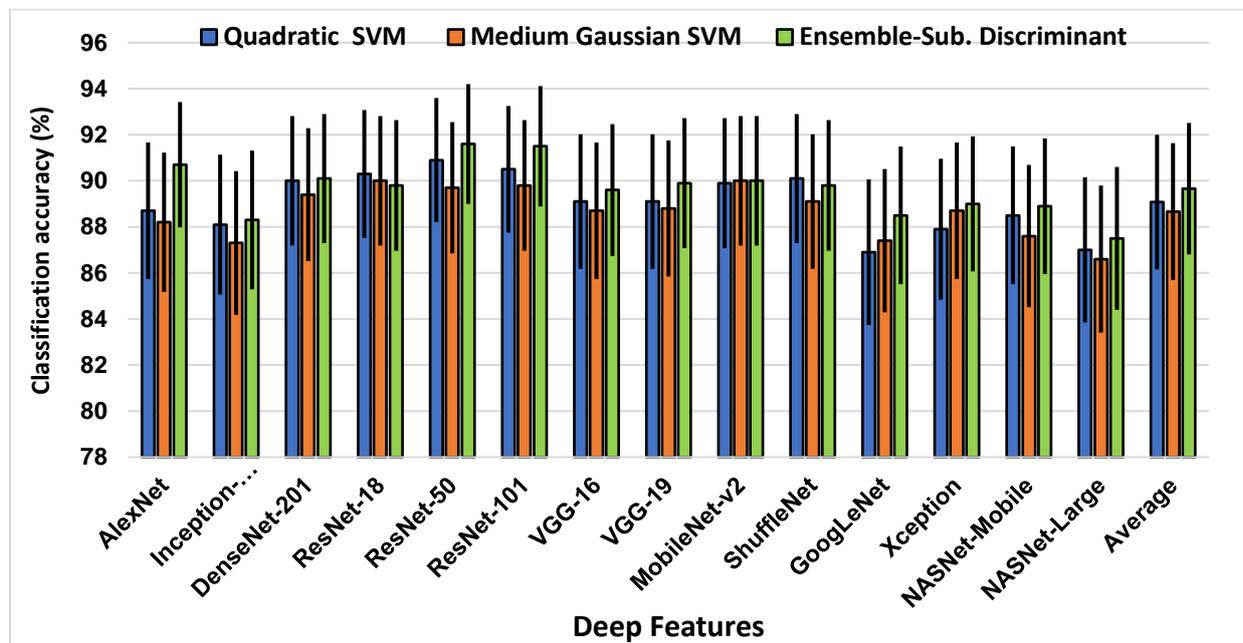

**Fig. 3** Classification accuracy for DF extracted from the 14 deep networks investigated in this study with 5-class COVID-19 images with 3 machine learning classifiers. The average of all classifiers is also shown with error bars representing standard deviation. Whiskers represent Confidence Interval (CI) calculated at the 95% confidence level of the detection accuracy.

The confusion matrix for the 5-class COVID-19 detection dataset with ResNet-50 and ensemble of subspace discriminant classifier is shown in Fig. 4. COVID-19 and TB achieved the best performance compared to the other classes, where less than 1% of the COVID-19 cases were mis-detected as normal and TB (Fig. 4), while the lowest performance was for pneumonia bacterial and viral. In addition, classification measures for each class in terms of precision, recall, specificity, *F*-score and AUC were also calculated (Table 4). The highest specificity was for COVID-19, normal and TB classes, which shows the potential use of our proposed method to aid for quick COVID-19 and TB detection.

**Table 3** The time needed to extract deep features from all pretrained networks for all 2186 images. Time estimated for 5-fold CV on Core *i5* CPU computer with 16 GB RAM.

| No. | Network Name | Time for deep feature extraction (minutes) for 2186 images |
|---|---|---|
| 1 | AlexNet | 3.25 |
| 2 | Inception-ResNet-v2 | 33.54 |
| 3 | DenseNet-201 | 20.79 |
| 4 | ResNet-18 | 6.50 |
| 5 | ResNet-50 | 6.90 |
| 6 | ResNet-101 | 11.60 |
| 7 | VGG-16 | 31.85 |
| 8 | VGG-19 | 38.72 |
| 9 | MobileNet-v2 | 5.37 |
| 10 | ShuffleNet | 4.62 |
| 11 | GoogLeNet | 7.16 |
| 12 | Xception | 24.75 |
| 13 | NASNet-Mobile | 8.07 |
| 14 | NASNet-Large | 94.05 |

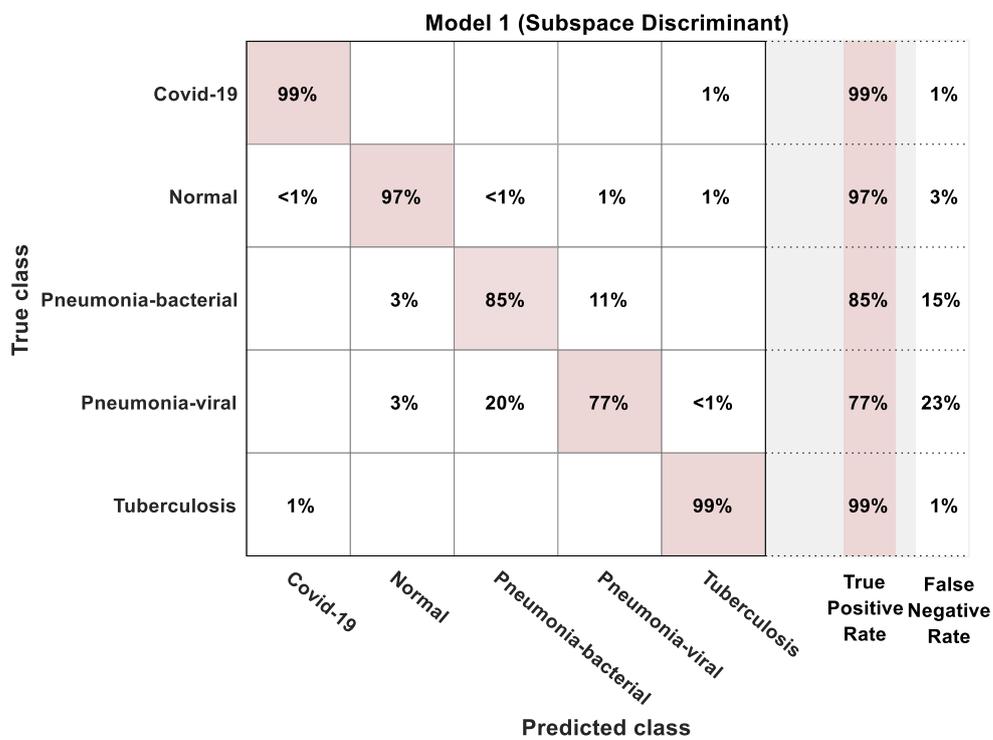

**Fig. 4** Confusion matrix for ResNet-50 for 5-class COVID-19 classification with ensemble of subspace discriminant classifier (5-fold CV).

Table 4 The classification performance including precision, recall, specificity, F-score and AUC for ResNet-50 and ensemble of subspace discriminant classifier. The overall detection accuracy is equal to 91.6 ±2.6 % (accuracy ± Confidence Interval (CI) at 95% confidence level)

| Class | Precision | Recall | Specificity | F-score | AUC |
|---|---|---|---|---|---|
| **COVID-19** | 99 | 98.6 | 99.8 | 98.8 | 1 |
| **Normal** | 94 | 97.2 | 98.5 | 95.6 | 0.99 |
| **Pneumonia-bacterial** | 81 | 85.4 | 95 | 83.1 | 0.97 |
| **Pneumonia-viral** | 86.3 | 77.2 | 97 | 81.5 | 0.97 |
| **Tuberculosis** | 97.3 | 99.3 | 99.3 | 98.3 | 1 |

The second part of the analysis was to investigate the class separability with the state-of-the-art high dimensional t-SNE visualization to understand how the features are distributed in the feature space (Fig.5). Clear separation of the COVID-19 and TB classes compared to that of normal can be observed, while there is an area of overlap of bacterial and viral pneumonia. Silhouette criterion values illustrated in Table 5, where the higher the values, the better the class separation, showed that COVID-19 and normal have a high Silhouette criterion values whereas pneumonia-bacterial has the lowest value (Table 5).

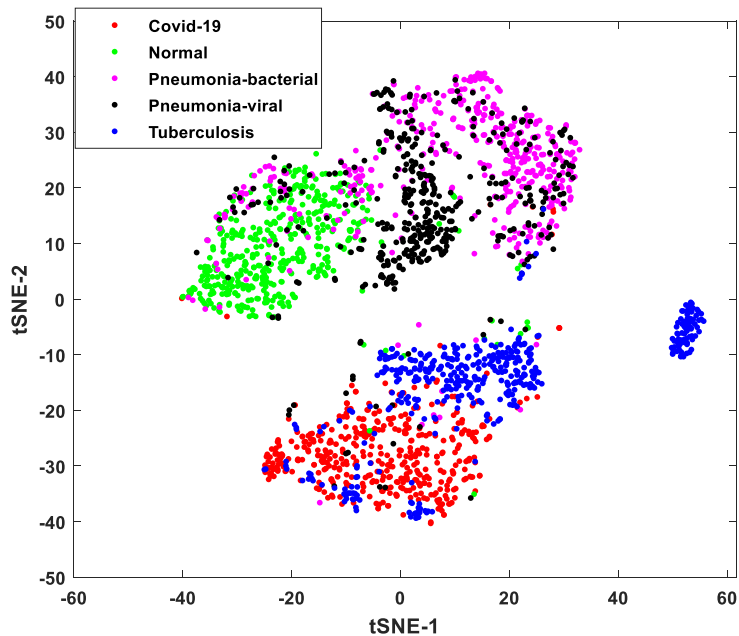

**Fig. 5** t-SNE visualizations of five-class balanced COVID-19 features extracted from ResNet-50 for 2186 X-ray images. See corresponding confusion matrix in Fig.4.

**Table 5** Silhouette criterion values for 5-class COVID-19 balanced dataset.

| Class | Silhouette criterion values |
|---|---|
| COVID-19 | 0.5345 |
| Normal | 0.5843 |
| Pneumonia-bacterial | -0.7114 |
| Pneumonia-viral | 0.3220 |
| Tuberculosis | -0.3190 |

Since there is an area of overlap between the 2-class of COVID-19 and TB to that of normal, we investigated three-class of COVID-19 vs. normal vs. TB due to the relevance of TB to some low- and middle-income countries [5]. Table 6 shows classification performance matrices for the three-class problem and the corresponding confusion matrix is illustrated in Fig. 6. The overall accuracy for the three-class problem was 98.6±1.4 % (accuracy ± CI). The t-SNE visualization displayed in Fig. 7 shows clear area of separation between normal and COVID-19 and also TB. There are some TB cases that are overlapped with COVID-19 cases.

**Table 6** The classification performance for 3-class (COVID-19 vs. normal vs. TB) with ResNet-50 and ensemble of subspace discriminant with 5-fold CV. The overall detection accuracy is equal to 98.6 ± 1.4 % (accuracy ± CI at 95% confidence level)

| Class | Precision | Recall | Specificity | F-score | AUC |
|---|---|---|---|---|---|
| COVID-19 | 98.39 | 98.39 | 99.20 | 98.39 | 1 |
| Normal | 100 | 98.86 | 100 | 99.43 | 1 |
| Tuberculosis | 97.49 | 98.62 | 98.74 | 98.05 | 1 |

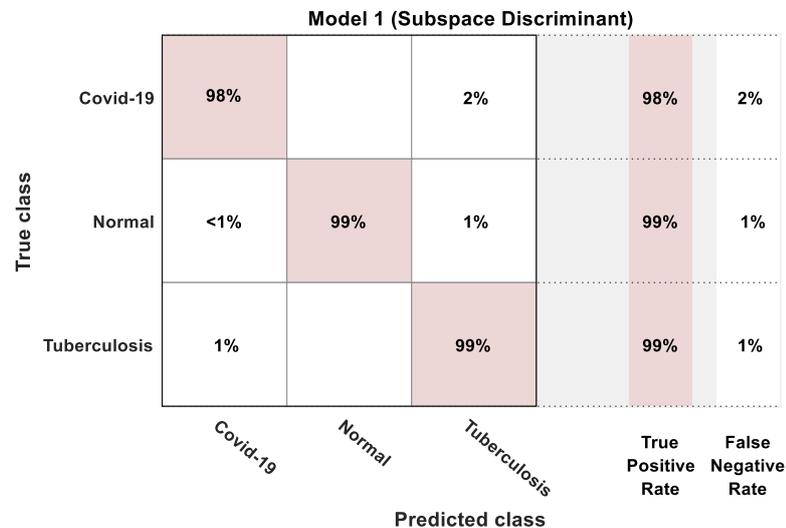

**Fig. 6** Confusion matrix for the 3-class (COVID-19 vs. normal vs. TB).

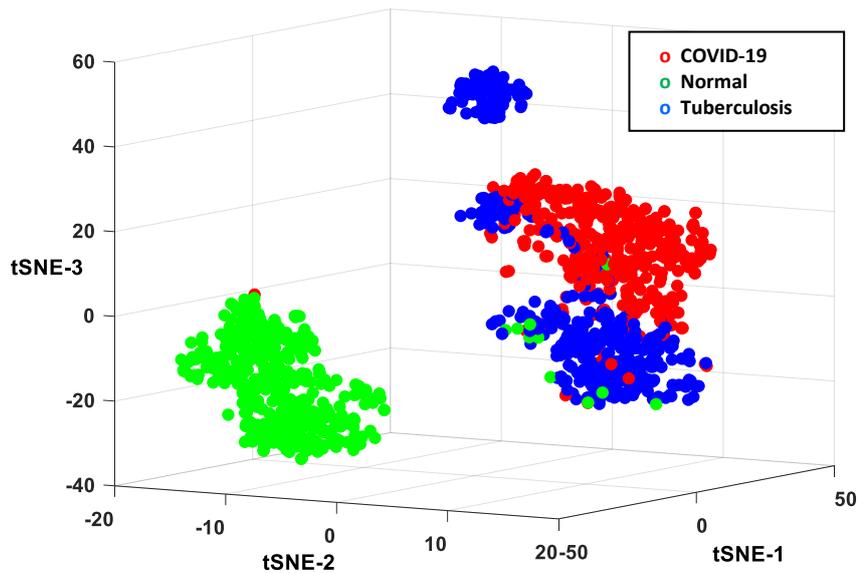

**Fig. 7** t-SNE visualization in 3 dimensions for the 3-class DF of ResNet-50 for COVID-19 vs. normal vs. TB. See corresponding confusion matrix in Fig.6.

Finally, to investigate the performance of our proposed COVID-19 detection method for 2-class COVID-19 vs. normal, similar to [4,11,12]. The results are shown in Fig.8 and t-SNE visualization is shown in Fig. 9. High detection accuracy rate of 99.9 was obtained on 847 images of COVID-19 and normal with five-fold CV.

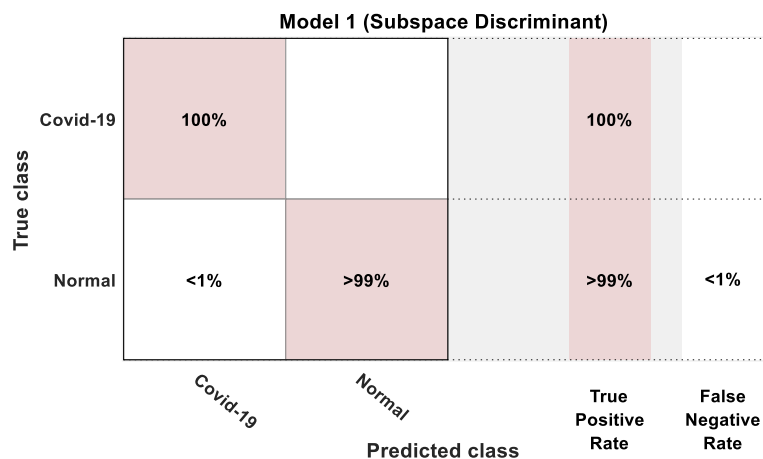

**Fig.8** Confusion matrix for 2-class of COVID-19 vs. normal. The detection accuracy is equal to 99.9 ± 0.5 % (accuracy ± CI at 95% confidence level)

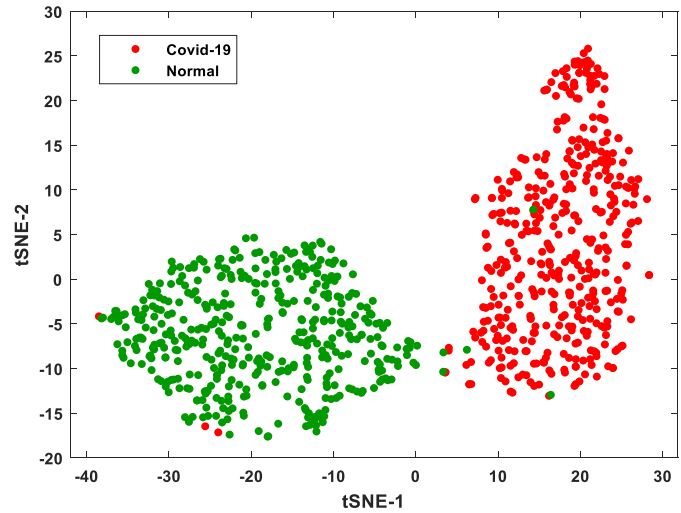

**Fig.9** 2-class t-SNE visualization for COVID-19 vs. normal. See corresponding confusion matrix in Fig.8.

We then ran a test with our proposed method on the four-class COVID-19 dataset [12] with four-fold CV for a straightforward comparison with that study. Our detection accuracy was equal to 90.2% which outperformed their accuracy (89.6%) [12]. Notably, our approach only required a CPU computer while their work needed a Tesla GPU machine.

**Model 1 (Subspace Discriminant)**

| True class \ Predicted class | Covid | Normal | Pneumonia-bacterial | Pneumonia-viral | True Positive Rate | False Negative Rate |
|---|---|---|---|---|---|---|
| Covid | 98% | <1% | 1% | 1% | 98% | 2% |
| Normal | <1% | 96% | 1% | 2% | 96% | 4% |
| Pneumonia-bacterial |  | 3% | 85% | 12% | 85% | 15% |
| Pneumonia-viral |  | 3% | 14% | 82% | 82% | 18% |

**Fig. 10** The confusion matrix of our proposed method on the four-class dataset used in [12].

## 4. Discussion

We presented in this paper an efficient approach employing existing and pretrained DL models for the classification of five-class COVID-19 images with three traditional machine learning classifiers. The main power of the features produced by these models is attributed to the fact that these networks learned the most discriminant parameters to classify the original images, upon which these models were developed, without any hand crafting feature extraction algorithms. Interestingly, the DF features were extracted from models not specifically trained on COVID-19 images but on general images of nearly 1000 classes and each model learned its own features or descriptors that best separate between the considered classes of images.

When observing the accuracy of the utilized DL models, it was found that ResNet-50 [46] had the highest classification accuracy across the quadratic SVM and the ensemble of subspace discriminant classifier, while being close to the best model results when using the medium Gaussian SVM classifier. Our $n$-way ANOVA analysis supported the statistical analysis of these results indicating a significant difference between the performance of the classification models across the 14 utilized DL models, with significant differences $\ll 0.001$ for all tests.

Once the results were statistically verified, we then studied the computational tie requirements needed to extract deep features from all pretrained 14 networks for 2186 images, with the results indicating that AlexNet took only 3.25 minutes on average across several runs to extract the important features from 2186 images. ResNet-50, being the best performing model on the current set of images, took approximately 0.19 seconds per image, run on standard Core $i$5 CPU computer (6.90 minutes for the whole dataset of 2186 images, Table 3). On the other hand, the worst performing model in terms of the computational cost was the NASNet-Large as it took around 94 minutes to extract the features. Other models had their computational costs ranging across their values. This in turn supports the first finding regarding the best performing model as ResNet-50 took only around a quarter of the average time for all models considered in Table 3.

COVID-19 and TB are two classes that are best detected among the five classes investigated in this study, specificity is equal to 99.8 % and 99.3%, respectively (Table 4). This may be clinically important, especially for some low and middle income countries, to detect both TB and COVID-19 since TB has a higher mortality rate more than COVID-19. The proposed pipeline can run on low resource settings (CPU computer instead of GPU).

The silhouette criterion values for the TB cases were low (Table 5), despite having a good detection accuracy (Table 4). This may be attributed that TB has two distinct clusters (Fig.5 and Fig.7), which increase the distance between the samples belonging to TB class, despite being distinct from normal and other types of pneumonia.

The subsequent analysis focused on ResNet-50, as we decided to further understand the performance of this DL network across each of the individual classes in the five-class dataset considered here. The importance of this step in the analysis can be appreciated when considering the similarity between some of the symptoms of the considered diseases, specially fever, tiredness and cough, for TB and COVID-19. Interestingly, the trained ResNet-50 model showed only 1% errors when it comes to the proper identification of COVID-19 and TB, while most of the confusion took a place between pneumonia bacterial and viral, with pneumonia viral being the most confused about class of diseases. These results were further supported by all the precision, recall, specificity, *F*-score and AUC measures for ResNet-50. These measures support the potential use of the proposed model as a quick and reliable CAD in COVID-19 and TB detection.

When inspecting the t-SNE visualizations of five-classes balanced COVID-19 features extracted from ResNet-50 for 2186 X-ray images, it was found that COVID-19 and TB were completely separated from normal and Pneumonia bacterial and viral. However, there was some degree of overlapping between COVID-19 and TB when visually inspecting the t-SNE feature projections across the first two dimensions. However, this visual overlapping was tackled with the power of our nonlinear classification models, in a 3D t-SNE visualization.

We also investigated the classification performance for three-class (COVID-19 vs. normal vs. TB) with ResNet-50 and ensemble of subspace discriminant and the two-class (COVID-19 vs. normal). In both set of tests, COVID-19 is shown to be completely separated from normal, and clearly separated from TB.

As we quantified the performance of our proposed combination of DL and machine learning models, we then analysed the reported performance results from several other research groups as shown in Table7 in comparison to our model results. Our proposed COVID-19 detection pipeline outperformed the state-of-the-art-literature, illustrated in Table 7. It should be noted that each model tested a different dataset, different testing split and also different classes. Hence, the comparison between these models here is for illustrative purposes but it shows the potential of our approach combining DF and simple classifiers in this task.

**Table 7.** Comparison to the previous literature on COVID-19 detection with X-ray images.

| Study | Dataset | Evaluation method | Techniques used | Detection accuracy |
|---|---|---|---|---|
| Narin et al. [26] | **2-class:** 50 COVID-19/ 50 normal | 5-fold CV | Transfer learning with Resnet50 and InceptionV3 | 98% |
| Panwar et al. [30] | **2-class:** 142 COVID-19/ 142 normal | Holdout 30% | nCOVnet CNN | 88% |
| Altan et al. [56] | **3-class:** 219 COVID-19 1341 normal 1345 pneumonia viral | Holdout 27% | 2D curvelet transform, chaotic salp swarm algorithm (CSSA), EfficientNet-B0 | 99% |
| Chowdhury et al. [11] | **3-class:** 423 COVID-19 1579 normal 1485 pneumonia viral | 5-fold CV | Transfer learning with ChexNet | 97.7% |
| Wang and Wong [21] | **3-class:** 358 COVID-19/ 5538 normal/ 8066 pneumonia | Holdout 30% | COVID-Net | 93.3% |
| Kumar et al. [36] | **3-class:** 62 COVID-19/1341 normal/ 1345 pneumonia | Holdout 30% | Resnet1523 features and XGBoost classifier | 90% |
| Sethy and Behera [35] | **3-class:** 127 COVID-19/ 127 normal/127 pneumonia | Holdout 20% | Resnet50 features and SVM | 95.33% |
| Ozturk et al. [4] | **3-class:** 125 COVID-19/ 500 normal 500 pneumonia | 5-fold CV | DarkCovidNet CNN | 87.2% |
| Khan et al. [12] | **4-class:** 284 COVID-19/ 310 normal/ 330 pneumonia bacterial/ 327 pneumonia viral | 4-fold CV | CoroNet CNN | 89.6%% |
| **This study** | **5-class:** 435 COVID-19/439 normal/ 439 pneumonia bacterial/ 439 pneumonia viral/ 434 Tuberculosis | 5-fold CV | Resnet50 features and ensemble of subspace discriminant classifier | 91.6 ±2.6%* |
|  | **3-class:** 435 COVID-19/439 normal/ 434 Tuberculosis |  |  | 98.6 ± 1.4%* |

\* Detection accuracy (Acc.) with 95% Confidence Interval (CI)
\*\* The CI for the previous literature is not provided.

The study has a potential limitation of relatively small number of COVID-19 and TB images, despite being the largest compared to previous literature so far. More COVID-19 and TB images are needed to improve the robustness of the proposed in a future research. The separation of the TB cases in two clusters in the t-SNE visualization should also be investigated in the future.

## 5. Conclusion

We proposed a combination of deep and machine learning models for the classification of X-ray chest images into five classes including COVID-19, in contrast with viral pneumonia, bacterial pneumonia, and TB, and in comparison, to the normal healthy subjects. Our work was motivated by the challenging conditions in low resource environments, where TB and COVID19 may be major healthcare problems. A key characteristic of our study is that the pre-trained networks can be utilized without GPU support as no (re-)training of the deep networks is required. Thus, we are able to extract DF of the X-ray images efficiently with a CPU-enabled computer. Our developments were tested using five-fold cross-validation, achieving $91.6 \pm 2.6$ % accuracy(accuracy $\pm$ CI) with a pipeline consisting of ResNet-50 for DF computation and ensemble of subspace discriminant classifiers in the distinction of the five groups or classes of diseases. In addition, we explored the classification accuracy in three- and two-class problems (COVID-19, TB and healthy cases; and COVID-19 and healthy cases; respectively) and obtained accuracies over 98% on both cases. Our study is limited by the relatively small number of COVID-19 and TB images, but it shows the promise of a pipeline requiring low computational resources to contribute to the detection of COVID-19 and TB using X-ray images where other more advanced or computationally demanding techniques may not be available.

**Declaration of Competing Interests**

None

**Acknowledgments**

The authors are grateful for the research groups who provided the X-ray images. We would like also to thank all medial staff around the world who are in the first line of defense against COVID-19. We are grateful for research community around the world who made research papers freely available for immediate download to facilitate worldwide research sharing to help in the current COVID-19 crisis.